\documentclass[apj]{emulateapj}
\newcommand{\DDir}{\relax{D\kern-.7em{/}}}

\newcommand{\be}{\begin{equation}}
\newcommand{\ee}{\end{equation}}
\newcommand{\bea}{\begin{equation*}}
\newcommand{\eea}{\end{equation*}}

\newcommand{\nin}{\relax{\in\kern-.8em{/}}}

\newcommand{\Gyr}{\mbox{ Gyr}}

\newcommand{\GHz}{\mbox{ GHz}}

\newcommand{\GeV}{\mbox{ GeV}}

\newcommand{\muG}{\mbox{ $\mu$G}}

\usepackage{amsmath}

\begin{document}

\title{Magnetic fields and cosmic rays in clusters of galaxies}

\author{Doron Kushnir\altaffilmark{1}, Boaz Katz\altaffilmark{1} and
Eli Waxman\altaffilmark{1}}
\altaffiltext{1}{Physics Faculty,
Weizmann Institute of Science, Rehovot, Israel}

\begin{abstract}
We argue that the observed correlation between the radio luminosity and the thermal X-ray luminosity of radio emitting galaxy clusters implies that the radio emission is due to secondary electrons that are produced by p-p interactions and lose their energy by emitting synchrotron radiation in a strong magnetic field, $B>(8\pi aT_{CMB}^4)^{1/2}\simeq3\muG$. We construct a simple model that naturally explains the correlation, and show that the observations provide stringent constraints on cluster magnetic fields and cosmic rays (CRs): Within the cores of clusters, the ratio $\beta_{\textrm{core}}$ between the CR energy (per logarithmic particle energy interval) and the thermal energy is $\beta_{\textrm{core}}\sim 2\cdot10^{-4}$; The source of these CRs is most likely the cluster accretion shock, which is inferred to deposit in CRs $\sim0.1$ of the thermal energy it generates; The diffusion time of 100~GeV CRs over scales $\gtrsim100$~kpc is not short compared to the Hubble time; Cluster magnetic fields are enhanced by mergers to $\gtrsim 1\%$ of equipartition, and decay (to $<1\muG$) on $1\Gyr$ time scales. The inferred value of $\beta_{\textrm{core}}$ implies that high energy gamma-ray emission from secondaries at cluster cores will be difficult to detect with existing and planned instruments.

\end{abstract}

% -------------------------- End of abstract -----------------------


% -----------------------------------------------------------------------
% --------------------------  Sec 1: INTRODUCTION -----------------------
% -----------------------------------------------------------------------

\section{Introduction}
\label{sec:Introduction}

Nonthermal emission is observed in several clusters of galaxies, mainly in the radio band: giant radio halos
(RHs) and mini-radio halos, fairly symmetric sources at the cluster center, as well as radio relics, elongated sources at the cluster periphery, have been observed \citep[e.g.,][]{feretti2008gcr}. The radio emission is interpreted as synchrotron radiation, thereby suggesting that relativistic electrons and magnetic fields are present in the intra-cluster medium (ICM). While the radio relics' emission can be attributed to merger or accretion shock waves \citep[e.g.,][]{ensslin1998crr}, the origin of the radio halos is not yet understood.

Pointed radio observations of a complete sample of about $50$ X-ray clusters at redshift $z = 0.2 - 0.4$, with rest-frame $[0.1,2.4]$~keV luminosity $L_{X}[0.1,2.4]\geq5\cdot10^{44}\,\textrm{erg}\,\textrm{s}^{-1}$, have been recently carried out \citep{venturi2007gmrt,brunetti2007crr,venturi2008gmrt}. One of the findings of these observations is that $\sim 30 \%$ of the X-ray luminous clusters host RHs, with all RHs being in merging clusters. In addition, a tight correlation between the radio luminosity and the X-ray luminosity of clusters with RHs was established. Note, that only a few halos have good multi-frequency observations, that allow one to estimate their integrated spectrum. These spectra are usually consistent with a spectral index of 1, $L_{\nu}\propto\nu^{-1}$ \citep{feretti2008gcr}.

Several models for the synchrotron emission in RHs have been presented in the literature. These models differ in the assumptions regarding the origin of the emitting electrons. In some models the emitting electrons are secondary electrons and positrons that were generated by p-p interactions of a CR proton population with the ICM \citep[e.g.][]{dennison1980frh,blasi1999crr} while in others the emitting electrons are reacelerated by turbulence from a preexisting population of nonthermal seeds in the ICM \citep[secondary or otherwise, e.g.][]{brunetti2001prc,petrosian2001one,brunetti2005arr,cassano2005cmn,cassano2007nsr,brunetti2008gre}.

In this paper we present a simple analytic model, following \citet[][]{kushnir2009non}, for the radio emission in clusters with RHs, and show that it naturally reproduces the observed correlation between the radio luminosity and the thermal X-ray luminosity in such clusters. The observations are described in \S~\ref{sec:observations}, and the simple model for the X-ray and radio emission from clusters with RHs is described in \S~\ref{sec:simple}. In this model, the radio emission is produced by secondary electrons, that are produced via p-p interactions of intra-cluster CRs with the ICM, and lose their energy by emitting synchrotron radiation in a strong magnetic field. It is assumed that the magnetic field is strong enough to ensure that synchrotron losses dominate the electrons' energy loss, and that the loss time is short compare to the cluster dynamical time. The validity of these assumptions as well as the implications of the bimodality of the radio luminosity distribution are discussed in \S~\ref{sec:evolution}. In \S~\ref{sec:models} we show that alternative models for the radio emission, with lower values of the magnetic field or different sources for the emitting electrons, are unlikely to reproduce the observed correlation between X-ray and radio luminosity. Our results are summarized and their implications are discussed in \S~\ref{sec:conclusions}.

We note that in some clusters nonthermal emission is also observed in hard ($>20\,\textrm{keV}$) X-rays (HXR) \citep[for review, see][]{rephaeli2008npc}. The HXR emission is usually interpreted as due to inverse Compton scattering of cosmic microwave background (CMB) photons by nonthermal relativistic electrons \citep[e.g.,][]{rephaeli1979rei,sarazin1999esp}. We discuss the origin of the nonthermal HXR emission in a separate paper \citep{kushnir2009hxr}. We show there that both the HXR observations and the results of the analysis presented in this paper, which indicates that $\beta_{\textrm{core}}\sim 2\cdot10^{-4}$, imply that the nonthermal HXR emission can not be produced by the same electrons producing the RHs. Both HXR and RH emission are consistent, instead, with a model where the nonthermal particles responsible for both radio and X-ray emission originate in cluster accretion shocks. In this model, the HXR emission is due to IC energy loss of electrons accelerated directly by the accretion shocks, while the radio emission is due to synchrotron emission of secondary electrons produced by the interaction of cosmic ray (CR) protons, which are also accelerated in the accretion shocks, with the thermal ICM protons \citep[see][for detailed discussion]{kushnir2009non,kushnir2009hxr}.

% -----------------------------------------------------------------------
% --------------------------  Sec 2: Observations -----------------------
% -----------------------------------------------------------------------

\section{Observations}\label{sec:observations}
\begin{figure}
\epsscale{1.2} \plotone{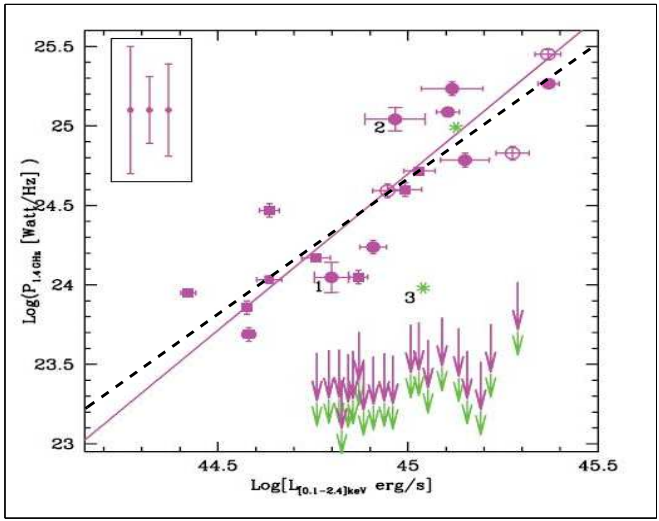} \caption{The distribution of clusters in the $P_{1.4}-L_{\textrm{X[0.1,2.4]}}$ plane, taken from \citet{brunetti2007crr} (reproduced by permission of the AAS). Over plotted (black dashed line) is the best linear fit, eq.~(\ref{eq:power law form}), for clusters that contain RHs (see text).
\label{Brunetti_fig}}
\end{figure}

We focus on the distribution of the clusters in the radio and X-ray luminosity plain. Pointed radio observations of a complete sample
of about $50$ X-ray luminous ($L_{X}[0.1,2.4]\geq5\cdot10^{44}\,\textrm{erg}\,\textrm{s}^{-1}$) clusters at redshift $z = 0.2 - 0.4$ have been
recently carried out \citep{venturi2007gmrt,brunetti2007crr,venturi2008gmrt}.
The distribution of clusters in the $P_{1.4}-L_{\textrm{X[0.1,2.4]}}$ plane is shown in figure~4 of \citet{brunetti2007crr}, where $P_{1.4}$ is the radio luminosity at $1.4\,\textrm{GHz}$. This figure is reproduced in figure~\ref{Brunetti_fig}.

The main characteristics of the distribution can be summarized as follows:
\begin{enumerate}

\item The radio luminosities of most of the clusters that contain RHs follow a linear correlation between $\ln(P_{1.4})$ and $\ln(L_{\textrm{X[0.1,2.4]}})$. The dispersion around the correlation for these clusters is relatively small: The standard deviation of $\ln(P_{1.4,\rm observed}/P_\text{1.4,expected})$ is $\approx\ln(1.7)$. Since this dispersion is much larger than the measurement errors, it reflects an intrinsic scatter of the radio luminosity (around the linear correlation).

\item The best linear fit, obtained assuming that the intrinsic variance around the correlation is $\Delta\ln(P_{1.4})=\ln(1.7)$, is\footnote{A somewhat steeper correlation, $P_{1.4}\propto L_{\textrm{X[0.1,2.4],45}}^{2}$, was obtained by \citet{brunetti2007crr}. This is shown in magenta in figure~\ref{Brunetti_fig}. It is not clear to us how this fit was
obtained.}
    \begin{eqnarray}\label{eq:power law form}
    &P_{1.4}\simeq 5\cdot10^{31}L_{\textrm{X[0.1,2.4],45}}^{1.7} \,\textrm{erg}\,\textrm{s}^{-1}\,\textrm{Hz}^{-1},
    \end{eqnarray}
    where $L_{\textrm{X[0.1,2.4],45}}=L_{\textrm{X[0.1,2.4]}}/10^{45}\,\textrm{erg}\,\textrm{s}^{-1}$.

\item About $\sim1/3$ of the clusters contain RHs. For bright clusters, $L_{X[0.1,2.4]}\gtrsim 10^{45}\,\textrm{erg}\,\textrm{s}^{-1}$, that do not contain a RH, the upper limit on $P_{1.4}$ is about an order of magnitude smaller than the value implied by the correlation.

\end{enumerate}

For a few clusters, we verified that the RH luminosities represent the total radio luminosities of the clusters by studying their radio surface brightness radial profiles \citep[taken from][]{cassano2007nsr,govoni2001rxr}.

As we show below, the observations described above allow us to derive stringent constraints on the magnetic fields and on the CR population within clusters.

% -----------------------------------------------------------------------
% --------------------------  Sec 3: Simple Model -----------------------
% -----------------------------------------------------------------------

\section{ A simple model for radio emitting clusters}\label{sec:simple}

In this section we present a simple model for the nonthermal emission of clusters that have RHs, and show that it naturally reproduces the observed correlation between the radio and X-ray luminosities.
In this model intra-cluster CRs interact with the thermal gas of the ICM to produce secondary electrons and positrons via p-p interactions. We assume that secondary electrons and positrons lose their energy primarily by emitting synchrotron radiation in a strong magnetic field, $B$, and that the energy loss time is short compared to timescales over which the magnetic field or the CR electron injection change considerably. Under these assumptions, the validity of which is discussed in \S~\ref{sec:evolution}, the radio synchrotron luminosity is equal to the energy generation rate of secondaries through p-p interactions. This generation rate is in turn correlated with the thermal Bremsstrahlung emission, as both processes (p-p and free free emission) are proportional to the gas density squared \citep{katz08}. As we show below, this implies a correlation between the radio luminosity and the X-ray luminosity, provided that the fraction of ICM thermal energy carried by CRs is roughly the same in different clusters.

In our simple model the ICM is composed of a thermal gas of ionized Hydrogen, with temperature $T$ and number density $n$, containing a population of proton CRs with a power law distribution $\varepsilon^{2}dn/d\varepsilon=\beta_{\textrm{core}}3nT/2$, and a magnetic field strong enough to make synchrotron emission the dominant energy loss process for the radio emitting electrons. Our choice of the CR spectral index, for which the energy density per logarithmic energy interval is independent of energy, is consistent with the observed spectral index of the radio emission, $L_\nu\propto\nu^{-1}$ \citep[see also a detailed discussion of the Coma cluster observations in][]{kushnir2009non}. The X-ray emission is produced in this model by thermal bremsstrahlung and has an emissivity (assuming the thermal Gaunt factor to be $1$ and neglecting heavier than Helium elements, which can increase the bremsstrahlung emissivity by a few tens of percent):
\begin{eqnarray}\label{eq:Lx app}
&\epsilon_{X}\approx f_{X}(\chi)\sigma_{T}\alpha_{e}c\sqrt{m_{e}c^{2}} T^{1/2}n^{2},
\end{eqnarray}
where $\sigma_{T}$ is the Thomson cross section, $\alpha_{e}$ is the fine structure constant and $f_{X}$ is a dimensionless factor of order unity which depends on the Hydrogen mass fraction $\chi$. The radio flux is produced by synchrotron emission of secondaries (electrons and positrons), which are the products of p-p collisions between the proton CRs and the thermal gas. We assume that the distribution of the radio emitting secondaries is in a steady state, where in the relevant energy bands the secondaries that are generated lose all their energy to synchrotron radiation. This implies that the synchrotron emissivity per logarithmic frequency interval is
\begin{eqnarray}\label{eq:sync nuenu}
\nu \epsilon_{\nu}^{\textrm{sync}}\approx
\frac{1}{2}\varepsilon \epsilon_{\varepsilon}^{e^{\pm}}.
\end{eqnarray}
Here $\varepsilon \epsilon_{\varepsilon}^{e^{\pm}}$ is the energy production rate of secondaries per logarithmic secondary energy interval, given by
\begin{eqnarray}\label{eq:pp app}
&\varepsilon \epsilon_{\varepsilon}^{e^{\pm}}\approx
0.1\cdot f_{\textrm{pp}}(\chi) \varepsilon^{2}\frac{dn}{d\varepsilon}n c\sigma_{\textrm{pp}}^{\textrm{inel}}.
\end{eqnarray}
$\sigma_{\textrm{pp}}^{\textrm{inel}}\approx40\,\textrm{mb}$ is the cross section for inelastic collision and $f_{\textrm{pp}}(\chi)$ is a dimensionless factor of order unity, which depends on the Hydrogen mass fraction.

Using equations~\eqref{eq:Lx app}, ~\eqref{eq:sync nuenu} and ~\eqref{eq:pp app} we see that in this model the radio luminosity per logarithmic frequency interval is proportional to the bolometric X-ray Luminosity,
\begin{eqnarray}\label{eq:radio to X}
\frac{\nu L_{\nu}^{\textrm{sync}}}{L_{X}}&\approx&
0.075\alpha_{e}^{-1}\beta_{\textrm{core}}\left(\frac{T}{m_{e}c^{2}}\right)^{1/2}\frac{ \sigma_{\textrm{pp}}^{\textrm{inel}}}{\sigma_{T}}\frac{ f_{\textrm{pp}}(\chi)}{f_{X}(\chi)} \nonumber\\
&\sim&1.1\cdot10^{-5}\beta_{\textrm{core},-4}T_{1}^{1/2},
\end{eqnarray}
where $\beta_{\textrm{core},-4}=\beta_{\textrm{core}}/10^{-4}$, $T_{1}=T/10\,\textrm{keV}$ and we used $\chi=0.75$. If $\beta_{\textrm{core}}$ does not vary considerably between clusters we expect a linear correlation between $\nu L_{\nu}^{\textrm{sync}}$ and $T^{1/2} L_{X}$.

\begin{figure}
\epsscale{1} \plotone{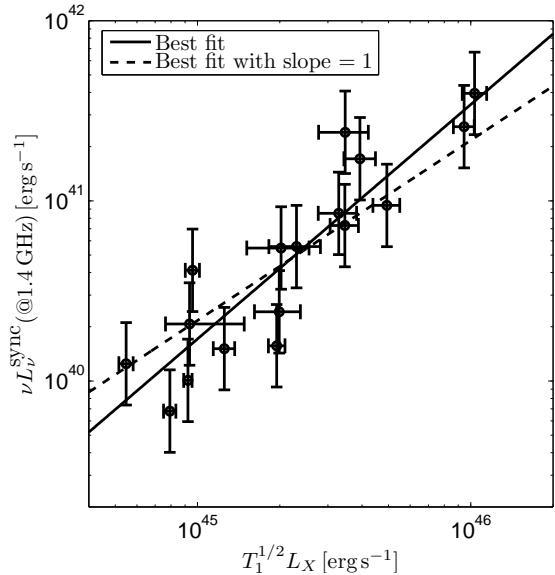} \caption{The distribution of the $17$ clusters of \citet{cassano2006sgr} in the $\ln(\nu L_{\nu}^{\textrm{sync}})$-$\ln(T^{1/2} L_{X})$ plane. The error bars of the radio luminosity describe our estimate of the intrinsic scatter, $\Delta\ln P=\ln(1.7)$. A linear fit for the correlation gives a slope of $1.3\pm0.7$ (best fit shown by the  solid line), and the best fit for a slope of $1$ (dashed line) gives $\beta_{\textrm{core},-4}\simeq2$ (see eq.~\ref{eq:radio to X}).
\label{P_TLbol}}
\end{figure}

Figure~\ref{P_TLbol} compares the prediction of eq.~(\ref{eq:radio to X}) with the distribution of the $17$ clusters of \citet{cassano2006sgr} in the $\ln(\nu L_{\nu}^{\textrm{sync}})$-$\ln(T^{1/2} L_{X})$ plane \citep[the values for $T$ and $L_{X}$, as well as the measurement errors in $T$ and $L_X$, are taken from][]{cassano2006sgr}. The error bars of the radio luminosity describe our estimate of the intrinsic scatter, $\Delta\ln P=\ln(1.7)$. A linear fit for the correlation between $\ln(\nu L_{\nu}^{\textrm{sync}})$ and $\ln(T^{1/2} L_{X})$ gives a slope of $1.3\pm0.7$, consistent with the predicted slope of 1. The best fit obtained for a slope of $1$ gives $\beta_{\textrm{core},-4}\simeq2$ (see eq.~(\ref{eq:radio to X})). The fact that there is a scatter of about a factor 2 around the correlation probably implies that there is a similar scatter in the value of $\beta_{\textrm{core}}$ among clusters. This is consistent with the scatter expected if the source of cluster CRs is acceleration in accretion shocks \citep[][see discussion in \S~\ref{sec:conclusions}]{kushnir2009non}.

Note that equation \eqref{eq:radio to X} can be applied to the ratio of the radio to X-ray surface brightness at any given point. For example, we give in table~\ref{tbl:cluster params} the central (bolometric) X-ray surface brightness, $F_{X}$, and the central radio surface brightness (taken as the maximal halo surface brightness), $\nu F_{\nu}^{\textrm{sync}}$, for the three clusters in \citet{govoni2001rxr}, for which such data are available. The value of $\beta_{\text{core}}$ derived from the data using equation~\eqref{eq:radio to X} is consistent with $\sim2\cdot10^{-4}$.

\begin{deluxetable*}{cccccrrrrrrrcrl}
\tablecaption{$\beta_{\text{core}}$ derived from surface brightness data \label{tbl:cluster params}} \tablewidth{0pt} \tablehead{ \colhead{Cluster name} &
\colhead{$F_{X}$$^{a,b}$} &
\colhead{$T$$^{a}$} &
\colhead{$\nu F_{\nu}^{\textrm{sync}}$$^{a,c}$} &
\colhead{$\beta_{\text{core,-4}}$$^{d}$} \\
\colhead{} & \colhead{[$\textrm{erg}\,\textrm{cm}^{-2}\,\textrm{s}^{-1}\,\textrm{arcmin}^{-2}$]} & \colhead{[$\textrm{keV}$]} & \colhead{[$\textrm{erg}\,\textrm{cm}^{-2}\,\textrm{s}^{-1}\,\textrm{arcmin}^{-2}$]} &  \colhead{} }
\startdata A520 & $1.6\cdot10^{-12}$ & 8.3 & $9\cdot10^{-17}$ & 5.7\\
A773 & $3.5\cdot10^{-12}$ & 8.6 & $5\cdot10^{-17}$ & 1.4\\
A2744 & $1.9\cdot10^{-12}$ & 10.1 & $5\cdot10^{-17}$ & 2.4\\
\enddata
\tablenotetext{a}{Data taken from \citet{govoni2001rxr}.}
\tablenotetext{b}{The central (bolometric) X-ray surface brightness.}
\tablenotetext{c}{The central radio surface brightness (taken as the maximal halo surface brightness).}
\tablenotetext{d}{Derived by using equation~\eqref{eq:radio to X}.}
\end{deluxetable*}

In order to compare the predicted correlation, eq.~(\ref{eq:radio to X}), with the correlation derived from the data shown in figure \ref{Brunetti_fig}, eq.~(\ref{eq:power law form}), we need to relate the bolometric X-ray luminosity $L_X$ to $T$ and to $L_{\textrm{X[0.1,2.4]}}$. $L_X$ and $L_{\textrm{X[0.1,2.4]}}$ are related by
\begin{equation}
L_{X}\approx L_{\textrm{X[0.1,2.4]}}\frac{2}{\int_{x_{\min}}^{x_{\max}}e^{-x/2}K_{0}(x/2)dx},
\end{equation}
where $x=h\nu/T$ and $K_{0}$ is the zeroth order modified Bessel function of the second kind \citep[we have included here the Gaunt factor corrections, which cannot be ignored at this level of approximation, see][]{drummond1961pp}. For the relevant cluster temperatures we may approximate
\begin{equation}\label{eq:TotToBand}
L_{X}\approx 3T_{1}^{0.6}L_{\textrm{X[0.1,2.4]}}.
\end{equation}
Next, it is well known that $L_X$ and $T$ are strongly correlated \citep[e.g.][]{markevitch1998lxt,arnaud1999lxt,reiprich2002mfx},
\begin{equation}\label{eq:LxTcorrelation}
L_{X}\approx L_{X0}T_{1}^{\alpha_{L}},
\end{equation}
with $\alpha_{L}$ in the range $2.5-3$ (The deviation of the luminosities from this phenomenological relation is remarkably small, of the order of tens of percents).
Using eq.~\eqref{eq:TotToBand} and eq.~\eqref{eq:LxTcorrelation}, eq.~(\ref{eq:radio to X}) gives
\begin{equation}
\nu L_{\nu}^{\textrm{sync}}\propto T^{1/2}L_{X}\propto L_{\textrm{X[0.1,2.4]}}^{1+\frac{1.1}{\alpha_L-0.6}}\,.
\end{equation}
Note, that the predicted slope of the correlation, $d\ln(\nu L_{\nu}^{\textrm{sync}})/d\ln(L_{\textrm{X[0.1,2.4]}})=1+\frac{1.1}{\alpha_L-0.6}$, depends weakly on the value of $\alpha_{L}$ (within the range of 2.5 to 3). Adopting $L_{X0}=3\cdot10^{45}$ and $\alpha_{L}=2.5$, our predicted correlation, eq.~(\ref{eq:radio to X}), may finally be written as
\begin{equation}\label{eq:P B gt Beq Lxb}
P_{1.4}\approx2.5\cdot10^{31}L_{\textrm{X[0.1,2.4],45}}^{1.6} \beta_{\textrm{core,-4}}\,\textrm{erg}\,\textrm{s}^{-1} \,\textrm{Hz}^{-1}.
\end{equation}
Comparing eq.~(\ref{eq:P B gt Beq Lxb}) with eq.~(\ref{eq:power law form}) we find that the predicted slope of the correlation is consistent with the observed slope, and that the normalization of the correlation is consistent with the observed one for $\beta_{\textrm{core}}\sim 2\cdot10^{-4}$.

Finally, let us determine the minimum magnetic field strength and the minimum time for variations of the magnetic field and of the electron injection, for which our model assumptions of fast electron cooling dominated by synchrotron losses are valid. The two main energy loss mechanisms of the electrons are synchrotron emission and inverse-Compton scattering of CMB photons. In order to ensure that the dominant process through which the electrons lose their energy is synchrotron radiation, the energy density in the magnetic field should be larger than the CMB energy density. This implies
\begin{eqnarray}\label{eq:B eq ucmb}
B>B_{\textrm{CMB}}\equiv(8\pi aT_{CMB}^4)^{1/2}
\approx3.2(1+z)^{2}\,\mu \textrm{G}.
\end{eqnarray}
This implies that the magnetic field energy caries a fraction
\begin{equation}\label{eq:epsilonB}
\epsilon_B\sim 1.7\cdot 10^{-2} T_1^{-1}n_{-3}^{-1}(B/B_{\textrm{CMB}})^{2}
\end{equation}
of the ICM thermal energy, where $n_{-3}=n/10^{-3}\,\textrm{cm}^{-3}$. $\epsilon_B$ values of order a few percent are reasonable, and observed in large-scale systems \citep{kim1989dir}. Note, that as long as $\epsilon_B\ll1$ the magnetic field does not affect the dynamics of the ICM.

The lower limit on $B$ sets an upper limit to the electron cooling time. The frequency at which an electron emits most of its synchrotron power is given by $\nu=\nu_{0}\gamma^{2}$, where $\nu_{0}=3eB/(4\pi m_{e}c)$ and $\gamma$ is the Lorentz factor of the electron. Thus,
\begin{eqnarray}
&\gamma_{\textrm{radio}}\approx 10^4\left(\frac{\nu_{\textrm{radio}}}{1.4\GHz}\right)^{1/2}\left(\frac{B}{B_{\textrm{CMB}}}\right)^{-1/2},
\end{eqnarray}
and the cooling time of the electrons is
\begin{eqnarray}\label{eq:t_cool}
&t_{\textrm{cool}}\approx 0.23\left(\frac{\nu_{\textrm{radio}}}{1.4\GHz}\right)^{-1/2}\left(\frac{B}{B_{\textrm{CMB}}}\right)^{-3/2}\Gyr.
\end{eqnarray}
For our model to be valid, the time scales for significant changes in the secondary injection and in the magnetic field must be larger than $\sim 0.25 (B/B_{\textrm{CMB}})^{-3/2}\Gyr$. An estimate for these time scales is given in \S~\ref{sec:evolution}, and it shown that they are much longer than the cooling time. Note, that since the energy of the secondary electrons is $\sim0.1$ of the energy of the parent CR protons leading to their production, and since the energy of CR protons in clusters is expected to reach $\sim10^{18}\,\textrm{eV}$ \citep[for details, see][]{kushnir2009non}, we expect the electron energy distribution to extend well above $\gamma\sim10^4$.

% -----------------------------------------------------------------------
% --------------------------  Sec 5: Field evolution -----------------------
% -----------------------------------------------------------------------
\section{Magnetic field evolution}\label{sec:evolution}

We next address the bimodality of the cluster radio luminosity distribution. Note, that since the cooling time of the CR protons at the relevant energies is very long compared to the Hubble time and their escape probability from the cluster is very low \citep[for details, see][]{kushnir2009non}, once CR protons are present within the cluster core in sufficient quantities to produce a RH, they remain there. Hence,
the absence of RHs in some clusters must be due, in our model, to lower values of the magnetic field in their ICM. Observations suggest that clusters that contain RHs show merger activity \citep{venturi2007gmrt}. This suggests the following simple scenario: ICM magnetic fields are enhanced by mergers to $B\gg B_{\textrm{CMB}}$, and later decay to lower values, $B\lesssim 1\muG$. Decay to $B\lesssim 1\muG=B_{\textrm{CMB}}/3$ implies a suppression of synchrotron luminosity by a factor of 10, rendering it undetectable. As explained in \S~\ref{sec:simple},     the magnetic field required to explain the radio luminosity, $B>B_{\textrm{CMB}}$, carries only a small fraction of the ICM thermal energy [eq.~\eqref{eq:epsilonB}]. Numerical simulations support the possibility of amplification of seed fields to such values by turbulence, generated in cluster mergers \citep[e.g.][]{ryu2008tmf}. We expect that in this scenario the magnetic fields will decay on time scales similar to the few Gyr time scales in which turbulence decays \citep[see e.g.][]{subramanian2006etm}.

Let us consider next the magnetic field decay time, which is required to account for the bimodality of the radio luminosity distribution. Before analyzing quantitatively the observed bimodality, two points should be stressed. First, only clusters from a complete sample can be used for a statistical study. We therefore restrict our analysis only to the complete sample presented in \citep{brunetti2007crr}. Note, that upper limits to the radio luminosity for clusters without RHs were obtained only for 20 of the clusters. Second, it is sensible to discus the bimodality in the $L_X-P_{1.4}$ plane only where the radio fluxes implied by the correlation are far above the detection limit (say a factor of 10). We thus restrict the discussion to X-ray luminosities $L_{X[0.1,2.4]}\gtrsim 10^{45}\,\textrm{erg}\,\textrm{s}^{-1}$.

In this restricted, complete sample there are 6 clusters for which the correlation holds, 1 cluster with a RH flux lower than implied by the correlation, and at least 9 clusters with radio luminosity that is smaller than the luminosity implied by the correlation by a factor of $>10$. These small numbers allow only a rough estimate of the decay time $t_{\text{dec}}$ of the magnetic field: Since the life time of massive clusters is $t_{\text{life}}\sim 5\Gyr$ and only $\sim1/3$ of the clusters have RHs, the typical decay time should be roughly $t_{\textrm{dec}}\sim t_{\text{life}}/3\sim 1.5\Gyr$.

A note is in order regarding the claim of \citet{brunetti2007crr, brunetti2008gre}, that the "empty" region in the $L_X-P_{1.4}$ plain (between the upper limits on $P_{1.4}$ for clusters with no RHs and the line defined by the $L_X-P_{1.4}$ correlation) implies that the magnetic fields must decay on $0.1-0.2\Gyr$ timescales. First, we note that
this statement was not supported by a quantitative statistical analysis. Moreover, the sample presented in \citep{brunetti2007crr} seems consistent with much longer decay times, of the order of few Gyr. Given the low statistics, the fact that there is at least 1 cluster with detected radio emission, which is smaller than implied by the correlation, is consistent with a scenario in which clusters spend similar times satisfying the correlation and producing radio flux lower than implied by the correlation (say $3\Gyr$ and $1.5\Gyr$ respectively), with $t_{\textrm{dec}}\sim 1 \Gyr$.

According to the scenario outlined above, the magnetic field is enhanced during mergers and later decays. In addition, an increase by a factor of order unity of the secondary injection rate (and X-ray luminosity) may take place during and following a merger, due to the increased densities and temperatures. These variations occur on time scales similar to, or larger than, the dynamical time scale of the mergers, $t_{\rm dyn}\sim1\Gyr$, which is much larger than the electron cooling time, given in eq.~(\ref{eq:t_cool}). This implies that our model assumption, that the electrons cool down on a time scale short compared to timescales over which the magnetic field or the CR electron injection change considerably, is valid.

It should be emphasized here that the possible increase in secondary injection rate following cluster mergers is not expected to modify the ratio of radio to X-ray luminosity given in eq.~(\ref{eq:radio to X}), and is not expected therefore to introduce a scatter to the predicted correlation, eq.~(\ref{eq:P B gt Beq Lxb}). This is due to the fact that fast cooling of the secondaries, $t_{\textrm{cool}}\ll t_{\rm dyn}$, ensures that they are in a quasi steady state for which  eq.~(\ref{eq:radio to X}) holds. Note that the ratio given in eq.~(\ref{eq:radio to X}) may be modified if significant particle acceleration takes place in merger shocks. However, such acceleration is not expected to significantly modify the existing CR population, due to the low Mach numbers of merger shocks \citep[see e.g.][]{ryu2003csw,gabici2003nrc,mccarthy2007msh,skillman2008csa,kushnir2009non}.

% -----------------------------------------------------------------------
% --------------------------  Sec 6: Other Models  -----------------------
% -----------------------------------------------------------------------
\section{Other models}\label{sec:models}
In the previous sections we presented a simple model for clusters with RHs in which the radiation is generated by secondary electrons radiating in strong magnetic fields, $B>B_{\textrm{CMB}}$. In this section we consider alternative models, where either $B<B_{\textrm{CMB}}$ or the radiating electrons are not secondaries produced by p-p collisions, and argue that such alternative models are unlikely to reproduce the observed correlation of radio and X-ray luminosities.

In models where $B<B_{\textrm{CMB}}$ \citep[e.g][]{dennison1980frh,blasi1999crr,brunetti2001prc,petrosian2001one,cassano2005cmn,cassano2007nsr} the synchrotron emission at radio frequencies strongly depends on the magnetic field value: $P_{\text{radio}}\propto B^{(p+1)/2}\propto B^{\alpha+1}$, where $p\approx 3$ and $\alpha\approx 1$ are the spectral indexes of the electron population and emitted radiation respectively, implying $P\propto B^2$. Any such model must thus assume a tight correlation between the magnetic field value $B$ and the X-ray luminosity, with an allowed deviation of a few tens of percents (to explain the small scatter in the correlation between the X-ray and radio luminosities). It is challenging to find a physical mechanism generating such a tight correlation between $B$ and $L_X$ \citep[see however][]{dolag2006sls}.
%Indeed, no such mechanism has been proposed in the $B<B_{\textrm{CMB}}$ models discussed in the literature. Instead, the magnetic field is chosen in such models in an ad hoc manner to reproduce the observed radio luminosity.
It should be added that for $B<B_{\textrm{CMB}}$ the fraction of energy carried by the magnetic field is small, $\epsilon_B=B^2/(8\pi)/(3/2nT)\sim 1.7\times 10^{-3} n_{-3}^{-1}T_1^{-1}B_{-6}$, where $B_{-6}=B/\mu\textrm{G}$, and thus large variations are not constrained by energetic considerations.

Let us consider next $B>B_{\textrm{CMB}}$ models in which the radiating electrons are not of secondary origin. For $B\gtrsim  B_{\textrm{CMB}}$, the radio luminosity and the rate of production of radiating electrons are related by $\varepsilon \epsilon_{\varepsilon}^e\approx 2\nu \epsilon_{\nu}^{\textrm{sync}}$ (assuming that the time scale for variations in $\varepsilon \epsilon_{\varepsilon}^{e}$ is longer than the cooling time). Since the synchrotron emission of secondary electrons accounts for the observed radio luminosity for $\beta_{\textrm{core}}\simeq10^{-4}$, the generation rate of high energy electrons required to account for the radio emission is given by (see eq.~\ref{eq:pp app})
\begin{eqnarray}\label{eq:e_rate}
&\varepsilon \epsilon_{\varepsilon}^{e}\approx
10^{-5}n^2T c\sigma_{\textrm{pp}}^{\textrm{inel}}.
\end{eqnarray}
In a model where the radiating electrons are secondaries from p-p collisions, the scaling $\varepsilon \epsilon_{\varepsilon}^{e}\propto n^2T$ is natural, and the normalization, $\varepsilon \epsilon_{\varepsilon}^{e}/n^2Tc\sigma_{\textrm{pp}}^{\textrm{inel}}\approx10^{-5}$ corresponding to $\beta_{\textrm{core}}\approx10^{-4}$, is naturally obtained for a reasonable value of the fraction of accretion shock energy converted to CRs (see discussion in \S~\ref{sec:conclusions}). It is challenging to account for the scaling and normalization of the high energy electron production rate in models where the radiating electrons are not secondaries.

Finally, we note that models, where the radiating electrons are not secondaries, require  $\beta_{\textrm{core}}\ll10^{-4}$ in order that secondary electrons would not dominate the radio flux. For example, this requirement is not satisfied in the model presented in \citep{brunetti2001prc,cassano2007nsr, brunetti2007ctg}, where the radiating electrons are not secondaries but rather preexisting nonthermal CR electrons reaccelerated by cluster turbulence. In this model
the CR density assumed in order to reproduce the observed radio luminosity (by emission from turbulence accelerated electrons) is a few percent, corresponding to $\beta_{\textrm{core}}\approx10^{-2}$. This implies that synchrotron emission from secondary electrons would dominate the radio luminosity in such models, and would exceed the observed radio luminosity by $\sim 2$ orders of magnitude.

% -----------------------------------------------------------------------
% --------------------------  Sec 7: Conclusions -----------------------
% -----------------------------------------------------------------------
\section{Discussion}\label{sec:conclusions}

We have presented (\S~\ref{sec:simple}) a simple model that explains the observed radio emission from clusters as synchrotron radiation from secondary electrons that are produced by p-p interactions and lose their energy by emitting radiation in a strong magnetic field, $B>(8\pi aT_{CMB}^4)^{1/2}\simeq3\muG$. The radio luminosity is given in this model by the rate of production of high energy electrons (eqs.~\ref{eq:sync nuenu}, \ref{eq:pp app}). This, combined with the observed correlation between the X-ray luminosity and the temperature of clusters (eq.~\ref{eq:LxTcorrelation}), naturally reproduces the observed radio and X-ray luminosity correlation, $P_{1.4}\propto L_{\textrm{X[0.1,2.4]}}^{1.7}$ (see eq.~\ref{eq:power law form}), provided that the fraction of ICM thermal energy carried by CRs per logarithmic particle energy interval is roughly the same for all clusters, and equal to $\beta_{\textrm{core}}\sim 2\cdot10^{-4}$ (see eq.~\ref{eq:P B gt Beq Lxb}).
We have shown (\S~\ref{sec:models}) that alternative models, where either $B<B_{\textrm{CMB}}$ or the radiating electrons are not secondaries produced by p-p collisions, are unlikely to reproduce the observed tight correlation of radio and X-ray luminosities.

The magnetic field implied by our model, $B\gtrsim B_{\textrm{CMB}}$, carries at least a few percents of the thermal energy of the ICM, $\epsilon_B\gtrsim10^{-2}$ [Eq.~\eqref{eq:epsilonB}]. This is much higher than the fraction $\beta_{\rm core, tot.}$ of thermal ICM energy carried by the CRs: for a $\varepsilon^2 dn/d\varepsilon\propto\varepsilon^0$ CR spectrum we have $\beta_{\rm core, tot.}\simeq10\beta_{\textrm{core}}\sim 10^{-3}\lesssim 0.1 \epsilon_B$. This is an
indication that these two nonthermal components are governed by
different processes (as we suggest below, the CR protons are
generated at the accretion shocks, while the magnetic fields are
amplified during mergers).

Clusters that contain RHs show merger activity \citep{venturi2007gmrt}. In order to explain the bimodality of clusters in the $P_{1.4}-L_{\textrm{X[0.1,2.4]}}$ plane reported by \citet{brunetti2007crr}, we suggested in \S~\ref{sec:evolution} that the magnetic field in clusters is amplified following mergers to $B\gg B_{\textrm{CMB}}$, and later decays on time scales of $1\Gyr$ to lower values, $B\lesssim 1\muG$. Numerical simulations support the possibility of amplification of seed fields to such values, which correspond to few percent of equipartition (eq.~\ref{eq:epsilonB}), by turbulence generated in cluster mergers \citep[e.g.][]{ryu2008tmf}. The time scale for magnetic field decay, implied by the $P_{1.4}-L_{\textrm{X[0.1,2.4]}}$ distribution, is similar to the few Gyr time scale of turbulence decay \citep[theoretical arguments supporting magentic field decay on a time scale similar to that of turbulence decay may be found, e.g., in][]{subramanian2006etm}. We emphasis that the observational fact that only $\sim1/3$ of the clusters host a RH (and these clusters show merger activity) is an indication
for deviation from the minimum energy state (for which the energy density in magnetic fields equals the CRs energy density), for clusters that host a
RH. Note, that for the other $\sim2/3$ of the clusters the minimum energy state
might hold, since in our scenario the CR protons' energy density is not
affected significantly by mergers, but the energy density of the
magnetic field decreases (after the merger) by a factor of $\sim10$ (at
least, to match the upper limits), which makes the energy densities of
the two nonthermal components roughly equal.

The observed tight correlation between $P_{1.4}$ and $L_X$ suggests that the fraction of ICM thermal energy carried by CRs is indeed uniform among clusters and given by $\beta_{\textrm{core}}\sim 2\cdot10^{-4}$. Various sources of CR were considered in the literature, including active galactic nuclei \citep[e.g.][]{katz1976oxr,fabian1976gcs,fujita2007nea,blasi2007grc}, dark matter bow shocks \citep[e.g.][]{bykov2000nec}, ram-pressure stripping of infalling galaxies \citep[e.g.][]{deplaa2006ces}, supernova (SN) explosions \citep[e.g.][]{volk1996nec} and shock waves associated with the process of large scale structure formation \citep[e.g.][]{loeb2000cgr,fujita2003nep,berrington2003npa,gabici2003nrc,brunetti2004arr,inoue2005hxr}.
The resemblance between cluster accretion shocks and collisionless non-relativistic shocks in the interstellar medium \citep{loeb2000cgr}, which are known to accelerate a power-law distribution of high energy particles, suggests that accretion shocks may be the source of cluster CRs \citep[Note that acceleration in merger shocks is not expected to significantly modify the existing CR population, due to the low Mach numbers of merger shocks, e.g.][]{ryu2003csw,gabici2003nrc,mccarthy2007msh,skillman2008csa,kushnir2009non}.

In order to account for the observed radio luminosity, the fraction $\eta_p$ of accretion shock energy deposited in relativistic CR protons should be a few percents. This conclusion is reached by noting that $\beta_{\rm core, tot.}$ is related to $\eta_p$ by $\eta_p\sim10\beta_{\rm core, tot.}$ \citep{pfrommer2007scr,kushnir2009non}. The latter relation is determined mainly by the reduction of the thermal energy fraction carried by CRs as the shocked gas is adiabatically compressed falling into the cluster's center. The compression of the gas leads to a reduction of the CR energy fraction, $\propto\rho^{-1/3}$, where $\rho$ is the gas density, due to the softer equation of state of the relativistic particles \citep[see][for a detailed discussion]{kushnir2009non}. The compression also increases the CR energy by a factor $\simeq10$, such that $\eta_p$ estimated above is relevant for $10\,\textrm{GeV}$ CR (note, that the parent CR energy is $\sim10$ times the energy of the secondary produced by an inelastic nuclear collision).

The accretion shock origin of cluster CRs is supported by both the inferred value of $\eta_p$ and the inferred scatter in $\beta_{\textrm{core}}$: The scatter in $\beta_{\rm core, tot.}$, in a model where CRs are accelerated in accretion shocks and advected towards the cluster center with the infalling gas, yields a scatter of roughly a factor of 2 in the value of $\beta_{\textrm{core}}$ among different clusters \citep{kushnir2009non}, consistent with the scatter implied by the $P_{1.4}-L_X$ relation; The implied acceleration efficiency of protons, $\eta_p\sim5$ percent, is similar to the acceleration efficiency of electrons in the accretion shocks, implied by the nonthermal HXR emission detected in several clusters \citep{kushnir2009non}.

We next discuss the possibility that the origin of ICM CRs is SN explosions \citep[e.g.][]{volk1996nec}. In this model, the observed Iron abundance of the ICM ($\varepsilon_{cl}\simeq1/3$ relative to the solar value $[Fe]_{\odot}$) is used to estimate the energy released by SN explosions, assuming some ratio between the energy released in SNe and the Iron mass ejected per explosion, $(E_{SN}/\delta M_{Fe})$. Assuming CRs are also injected in such explosions with acceleration efficiency per logarithmic energy bin of $\beta_{CR}^{SN}$, the ratio of the energy density of such CRs to the thermal energy density of the ICM per logarithmic energy bin is given by
\begin{eqnarray}\label{eq:CR SN}
\beta_{core}^{SN}\simeq\frac{\mu m_{p}}{(3/2)T}\varepsilon_{cl}[Fe]_{\odot}\frac{E_{SN}}{\delta M_{Fe}}\beta_{CR}^{SN}F_{\textrm{cool}}.
\end{eqnarray}
The last factor $F_{\textrm{cool}}$ represents the energy loss of these CRs as they leave galaxies into the ICM. The ratio between $\beta_{core}^{SN}$ and the corresponding value for accretion shocks $\beta_{core}^{\textrm{acc}}$ is given by
\begin{eqnarray}\label{eq:CR SN 2}
\frac{\beta_{core}^{SN}}{\beta_{core}^{\textrm{acc}}}&\simeq&1.6\frac{\beta_{CR}^{SN}}{\beta_{CR}^{\textrm{acc}}} F_{\textrm{cool}}\nonumber\\
&\times&\left(\frac{E_{SN}}{10^{51}\,\textrm{erg}}\right) \left(\frac{\delta M_{Fe}}{0.1M_{\odot}}\right)^{-1}T_{1}^{-1},
\end{eqnarray}
where $\beta_{CR}^{\textrm{acc}}$ is the acceleration efficiency per logarithmic energy bin in accretion shocks, and we included the reduction of the thermal energy fraction carried by CRs as the shocked gas is adiabatically compressed falling into the cluster's center. The adopted value for $(E_{SN}/\delta M_{Fe})$ is probably an upper limit, since it is derived by assuming that all of the Iron is generated in type II SNe \citep[see however][for the large uncertainty in this value]{hamuy2003opp}. However, some of this Iron can be produced in other processes, especially type Ia SNe with $\delta M_{Fe}\simeq0.7M_{\odot}$. Furthermore, $F_{\textrm{cool}}$ is probably a small number, such that if the acceleration efficiency in both SNe and accretion shocks is similar, than we expect accretion shock CRs to dominant those produce in SNe.

In our analysis we have so far neglected the possible effects of CR diffusion. If $100\,\textrm{GeV}$ CR diffusion is significant over scales $\gtrsim100\,\textrm{kpc}$, CRs may "escape" the infalling gas, possibly leading to $\beta_{\rm core, tot.}/\eta_p\ll1/20$. A discussion of the transport of CRs in the collisionless cluster plasma, a subject which is poorly understood, is beyond the scope of this paper \citep[note, that CR diffusion is poorly understood also in the Galaxy, where much more data are available on the CRs and on the magnetic field configuration, e.g.][]{ptuskin2006ern}. We note, however, that if diffusion were important, allowing most of the CRs accelerated in the accretion shock not to reach the cluster core, then $\beta_{\textrm{core}}$ would have been likely to vary considerably between clusters. The fact that we observe a tight correlation between the radio and X-ray luminosities indicates that this is not the case.

According to our model, the clusters' radio luminosity is given by the energy production rate of charged, $e^\pm$, secondaries. This implies that the high energy gamma-ray luminosity produced at the cluster core by the decay of neutral, $\pi^0$ secondaries, is proportional to the radio luminosity and should be given by ${\nu L_{\nu}^{\gamma-\textrm{ray}}\simeq3\cdot10^{41}L_{\textrm{X[0.1,2.4],45}}^{1.7}\,\textrm{erg}\,\textrm{s}^{-1}}$. The predicted core luminosity will be difficult to detect with existing and planned instruments. Note, however, that high energy inverse Compton gamma-ray emission, from electrons accelerated at the accretion shocks, should be detectable by FERMI and possibly by existing Cherenkov telescopes \citep[see detailed discussion in][]{kushnir2009non}.

Finally, a cautionary note should be made regarding predictions for low frequency radio observations of clusters (e.g. with ALMA, LOFAR). Under our model assumptions, the characteristic energies of electrons dominating the emission at low frequencies are $\simeq1.4(\nu_{\textrm{radio}}/100\,\textrm{MHz})^{1/2}(B/B_{\textrm{CMB}})^{-1/2}\,\GeV$, which are the products of CRs that were accelerated to $\sim1\,\textrm{GeV}$ energies. Since the spectrum of CRs is poorly understood at energies $\varepsilon\sim m_{p}c^{2}$, the extrapolation of the radio spectrum from the currently observed high frequencies to the relevant low frequencies is uncertain.

\acknowledgments The authors thank G. Brunetti, C. Pfrommer, A. Loeb and W. Hofmann for careful reading of the manuscript and for useful discussions. This research was partially supported by ISF, AEC and Minerva grants.

% -------------------------- End of Discussion --------------------------

%-----------------------------------------------------------------------------
% --------------------------      BIBLIOGRAPGHY ---------------------------
%-----------------------------------------------------------------------------

%\clearpage

\bibliography{ms1}
\bibliographystyle{hapj}

% ------------------------------ End of bibliography --------------------

\end{document}